# ESCAPE

Preparing Forecasting Systems for the Next generation of Supercomputers

# D4.5 Report on energy-efficiency evaluation of several NWP model configurations

Dissemination Level: Public

This project has received funding from the European Union's Horizon 2020 research and innovation programme under grant agreement No 67162

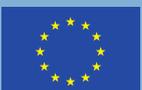
Funded by the European Union

Co-ordinated by

# ESCAPE

**Energy-efficient Scalable Algorithms for Weather Prediction at Exascale**

Author **Joris Van Bever (RMI), Alex McFaden (OSYS), Zbigniew Piotrowski (PSNC), Daan Degrauwe (RMI)**

Date **30/05/2018**

# Table of Contents

# 1 Executive Summary

In this deliverable we report on energy consumption measurements of a number of NWP models/dwarfs on the Intel E5-2697v4 processor. The chosen energy metrics and energy measurement methods are documented. Energy measurements are performed on the Bi-Fourier dwarf (BiFFT), the Acraneb dwarf, the ALARO 2.5 km Local Area Model reference configuration (Bénard et al. 2010, Bubnova et al. 1995) and on the COSMO-EULAG Local Area Model reference configuration (Piotrowski et al. 2018). The results show a U-shaped dependence of the consumed energy on the wall-clock time performance. This shape can be explained from the dependence of the average power of the compute nodes on the total number of cores used.

The relative energy consumption contributions of the BiFFT dwarf to the ALARO reference configuration show a different behavior compared to its relative contribution to the wall-clock time. While communications cause a significant increase to the relative wall-clock time with increasing number of cores, the relative energy consumption contribution actually decreases slightly. This is simply due to the fact that the energy consumption of the communications are not accounted for. The energy consumption pie charts are therefore more representative of the relative computational workloads of the dwarfs.

The comparison between the ALARO 2.5 km and COSMO-EULAG 2.2 km reference configuration indicates that the latter may be more energy consuming and requires slightly more runtime. However, a fair conclusion is not straightforward because of slightly differing setups (especially as far as the used timestep is concerned).

We compare the energy consumption of the BiFFT dwarf on the E5-2697v4 processor to that on the Optalysys optical processors. The latter are found to be much less energy costly, but at the same time it is also the only metric where they outperform the classical CPU. They are non-competitive as far as wall-clock time and especially numerical precision are concerned.

Finally, we suggest an interesting candidate energy metric called the *energy roofline*. It is an energy analogue to the well-known time-based performance roofline. It allows an assessment of whether a piece of code is compute-bound or memory-bound in terms of energy consumption. Since the dependent quantity in both roofline models is the same (i.e. arithmetic intensity), a straightforward assessment can be made of whether a piece of code has a balance gap, i.e. a difference between the arithmetic intensities whereby the code becomes compute-bound in terms of performance as compared to energy consumption. This also allows an estimate as to whether that piece of code is easier to optimize for performance or for energy efficiency. Such a tool could be useful for more detailed future energy optimization studies for NWP models/dwarfs.

## 2 Introduction

### 2.1 Background

ESCAPE stands for Energy-efficient Scalable Algorithms for Weather Prediction at Exascale. The project develops world-class, extreme-scale computing capabilities for European operational numerical weather prediction and future climate models. ESCAPE addresses the ETP4HPC Strategic Research Agenda 'Energy and resiliency' priority topic, promoting a holistic understanding of energy-efficiency for extreme-scale applications using heterogeneous architectures, accelerators and special compute units by:

- Defining and encapsulating the fundamental algorithmic building blocks underlying weather and climate computing;
- Combining cutting-edge research on algorithm development for use in extreme-scale, high-performance computing applications, minimizing time- and cost-to-solution;
- Synthesizing the complementary skills of leading weather forecasting consortia, university research, high-performance computing centers, and innovative hardware companies.

ESCAPE is funded by the European Commission's Horizon 2020 funding framework under the Future and Emerging Technologies - High-Performance Computing call for research and innovation actions issued in 2014.

### 2.2 Scope of this deliverable

#### 2.2.1 Objectives of this deliverable

This deliverable is the result of Task 4.4 in work package 4. Its goal is to propose energy-aware metrics and measurement methodology useful in a numerical weather prediction (NWP) context. These methodologies are to be applied to a selected number of individual dwarfs and reference configurations (see D4.3). The results of such energy consumption measurements on these dwarfs and reference configurations are reported and discussed in relation to their wall-clock times.

#### 2.2.2 Work performed in this deliverable

The planned work consisted of 4 subtasks:

1. Definition of energy metrics.
2. Evaluate energy of alternative advective transport options.
3. Evaluate energy efficiency of optical processors for the bi-Fourier transform in a local area model (LAM) context.

4. Test the energy-efficiency of different time-stepping strategies in a LAM context.

The performed work consists of:

- A description of the energy metrics used and the chosen methodology to measure energy.
- A report of the results of energy measurements of the ALARO 2.5 km reference configuration, the COSMO-EULAG reference configuration, the Bi-Fourier dwarf (BiFFT) and the ACRANEB2 radiation dwarf.
- A comparison between the runtime performance/energy consumption of the optical processors and the Intel E5-2697v4 processor when performing a Bi-Fourier transformation.

### 2.2.3 Deviations and counter measures

Initially, subtasks 2 to 4 were meant to be carried out with the help of the LAM extended version of `Atlas` (see. D4.4). Due to as yet unimplemented `Atlas` features involving numerical operators concerning geographical projections in the `Atlas` LAM version, subtasks 2 and 4 could not be performed.

Since Deliverable D3.3 already demonstrated the suboptimal numerical accuracy of the optical processors for NWP purposes, we chose not to make the time-consuming interface towards the ALARO reference configuration as planned for Subtask 3. Instead, we just compared energy measurements and runtime performance between both hardware components without an explicit coupling.

## 3 Energy measuring and metrics

Many energy metrics can be proposed, but here we will focus on total computational energy consumed and average computational power. Section 3.1 describes the method used to measure energy on the `cca` cluster at ECMWF. In section 3.2, we briefly mention an interesting general energy metric, which looks promising for code optimization purposes, called the energy roofline (Choi et al. 2013). It is built as an energy analogue to the classical time-based roofline model and might be a good candidate model for future studies in the energy profiling and optimization of NWP codes and dwarfs.

### 3.1 PAPI Cray Power Management (PM) counters

As described on the Cray Documentation Portal; the Cray XC series systems support two types of power management counters. The PAPI Cray RAPL component provides socket-level access to Intel Running Average Power Limit (RAPL) counters, while the similar PAPI Cray Power Management (PM) counters provide compute node-level access to additional power management counters. Together, these counters enable the user to monitor and report energy usage during program execution.

When RAPL counters are specified, one core per socket is tasked with collecting and recording the specified events. When PM counters are specified, one core per compute node is tasked with collecting and recording the specified events.

The update frequency of these counters is about 10 Hz, meaning that very short programs cannot be accurately measured. Furthermore, the counters do not allow the measurement of parts of a program (e.g. a subroutine or an individual loop).

We used these PM counters to perform the energy measurements reported below.

### 3.2 Energy roofline model

A promising general energy metric in terms of code optimization is the energy roofline see (Choi et al. 2013). As is the case for the classical time-based roofline model, this metric was not designed specifically for numerical weather prediction applications, but could nevertheless provide very useful insights in the energy consumption behaviour of any code. It builds on the concept of a time-based roofline to make an analogues energy roofline which can be readily compared to the former. The concepts discussed below are a short summary of Choi et al. (2013).

In time-based roofline models, a key quantity is the arithmetic or computational intensity; the ratio of work the system can perform per unit of data transfer. Analogously, one can define a quantity that describes the amount of work performed per unit of consumed energy. This quantity can then serve to develop an energy-based analogue of the time-based roofline model.

Figure 1 shows the difference between a time-based roofline and an energy roofline model. On the horizontal axis (which represents arithmetic intensity, $I$), both the time-balance point and the energy-balance point are marked. As is well known, the time-balance point ($B_\tau = 3.6$ flop/byte in figure 1) is the arithmetic intensity for which the rate of computational work equals the rate of memory access. It separates the memory-bound region to the left from the compute-bound region to the right. The energy-balance point ($B_\epsilon = 14$ flop/byte in figure 1) analogously represents the arithmetic intensity for which the computational energy consumption equals the memory access energy consumption and similarly separates energy-equivalent regions of compute boundedness from regions of memory-boundedness.

Contrary to the time-based roofline, the energy roofline has an arched 'roof' due to the fact that runtime can be overlapped while energy consumption cannot. An interesting consequence is that if time-balance differs from energy-balance, then there are distinct notions of being compute-bound versus memory-bound, depending on whether the optimization goal is to minimize time or to minimize energy. This difference can be measured as a time-energy balance gap.

As discussed in Choi et al. (2013), if constant power (i.e. the power needed to operate the system) is ignored, one expects $B_\tau < B_\epsilon$. In that case, an algorithm with $B_\tau < I < B_\epsilon$ is simultaneously compute-bound in time while being memory-bound in energy. If we also assume that increasing arithmetic intensity is the hard part about designing new algorithms or tuning code, then $B_\tau < B_\epsilon$ suggests that energy-efficiency is even harder to achieve than time-efficiency. The balance gap, or ratio $B_\epsilon/B_\tau$, can then be used as a measure of this difficulty.

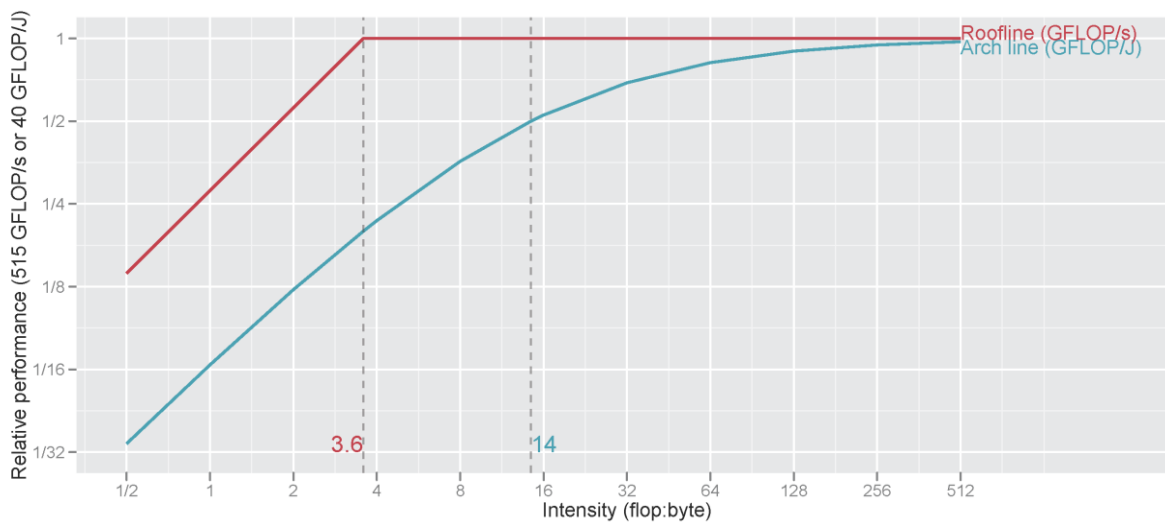

*Figure 1: Rooflines versus arch lines. The red line with the sharp inflection shows the roofline for speed; the smooth blue line shows the "arch line" for energy-efficiency. The time- and energy-balance points (3.6 and 14 FLOP/Byte, respectively) appear as vertical lines and visually demarcate the balance gap. (From Choi et al. (2013).)*

In contrast to the Cray PM counters which can only work for sufficiently long run-times due to their low sampling frequency, the energy roofline should be applied to individual parts of a code like a loop or subroutine, i.e. a part of the code for which an meaningful arithmetic intensity can be defined. For this to work, one requires profiling tools which can allow instrumentation of the code and which operate at much higher sampling frequencies. Unfortunately, such tools are not as yet available on `cca`. For the sake of this deliverable we therefore focused on energy versus wall-clock time using the tools described in the next section.

## 4 Energy measurements on `cca`; the Cray XC40 cluster *at ECMWF*

### 4.1 Method

On the `cca` cluster, we used the PAPI Cray Power Management (PM) counters to perform our energy measurements. These counters enable the user to monitor and report energy consumed. Access to the energy data of a compute node is obtained through a set of files located in the directory `/sys/cray/pm_counters/`. For our purposes, the relevant files are:

- `/sys/cray/pm_counters/energy`, and
- `/sys/cray/pm_counters/startup`.

The content of the energy file provides the accumulated energy in Joules of the compute node up to this point in time, relative to the last change of the counter inside the startup file. This means that to obtain the consumed energy of the compute node between two points in time, the content of the energy file is read twice and the earliest value is subtracted from the latest value. Note, however, that each energy reading must be accompanied by a reading of the startup file to make sure that both energy values are with respect to the same startup counter, otherwise they are meaningless and the job run must be made anew. The startup counter is changed when e.g. the power of the compute node exceeds a cap value.

On the basis of these counters, a script was implemented allowing for the simultaneous measurement of consumed energy as well as wall-clock time (in units of ms) for any set of bash commands (e.g. an executable). The script in question needs to be called right before the said set of commands as well as right after.

The `np` queue was used at all times, even for jobs using fractions of a node. This ensures that the entire node was dedicated to our job, preventing any contamination to the energy consumption results by non-related jobs. Note that the measurements only represent the consumption of energy due to the computational work of the compute nodes. We have no means of directly measuring any additional energy consumption due to e.g. communication or cooling of the cluster. On the other hand, it is possible to estimate the energy consumption of one compute node during an idle state. It was determined by measuring the energy consumption during a 1 minute sleep command. This helps to compare the computational energy consumption of the model codes to the energy needed to operate the node.

We have focused on the dwarfs and reference configurations as installed on the `cca` cluster at the ECMWF. The measurements therefore pertain to the performance of the Intel Xeon E5-2695v4 "Broadwell" processors.

### 4.2  BiFFT

The settings of the BiFFT dwarf were chosen for the purpose of also estimating the relative contribution of energy consumption of the BiFFT component in the ALARO reference configuration (see below). Besides the pie charts of wall-clock time contributions for each dwarf to the reference configuration shown in D4.3, this will allow us to make estimations of energy consumption contributions as well.

For the BiFFT dwarf, one direct and one inverse transformation are performed per loop iteration for the prescribed number of fields. In the ALARO reference configuration, the inverse transformations are performed on double the number of fields compared to the number of direct transformations; 350 direct ones versus 700 inverse ones. Therefore, to level the playing field we chose to set-up the BiFFT dwarf to perform 525 loop iterations, resulting in the same total number of spectral transformations performed compared to the ALARO reference configuration. As in the latter, 65 vertical levels were assumed in all cases. (Note that the chosen number of loop iterations also accounts for field derivatives which need to be transformed from spectral space to grid point space as well.) We have used full nodes only, except of course for the serial run.

Figure 2 shows the results for the energy consumption vs. wall-clock time for the BiFFT dwarf. It is clear that the energy consumption for all multi-core runs is dominated by the computations. However, the total energy consumption of the serial run, at an average power of 0.12 kW, does contain a significant contribution due to the operation of the node.

As already mentioned in D4.3 and later confirmed in D3.3, the wall-clock time of the BiFFT dwarf does not scale very well. This can be seen in the left part of Figure 2 where the wall-clock times of the many-node runs tend toward a vertical profile, i.e. ever increasing energy consumption without much gain in speed-up. There is clearly still room for optimization for the BiFFT dwarf. In its current state the optimal configuration seems to be in the range between 30-150 cores or so.

### 4.3   Acraneb

The version used for these measurements is the Acraneb2 radiation scheme as refactored for D1.2 to run optimally on 2016 Intel Xeon processors. The code of the Acraneb dwarf contains no MPI nor OMP parallelization directives, since this is implemented higher up in the stack of the ALADIN code. To nevertheless estimate the behavior of the dwarf for multi-node setups we simply assumed that the energy consumption of $n$ full nodes equals $n$ times that of 1 full node while the wall-clock time remains the same. Therefore each simulation consisted of 1 full node with an adapted workload which corresponds to what it would have been in a true parallel run. As with the BiFFT dwarf, the Acraneb dwarf was run with a 540 x 450 grid and 65 vertical levels to correspond to the ALARO runs (which are described in the next section). The same total of 525 spectral transformations were used as for the BiFFT case (see sect. 4.2).

The workload of 1 full node was computed in each case, depending on the total number of simulated MPI tasks. Since our previous wall-clock time measurements showed that the Acraneb dwarf scales rather well (see D4.3), the results obtained here (see Figure 3) should not differ significantly from true parallelization experiments (were they possible).

The graph (Figure 3) is similar to the BiFFT one except that here we have also included measurements on a partially filled node. Starting from the serial run at the top right, as the number of used cores increases initially the data points more or less follow a line of constant average power. This indicates that the average power used by a compute node does not depend strongly on the fraction of its used cores (see also Figure 5 in the next section). In that case, a good scaling executable will initially show both decreasing wall-clock times as well as consumed energy since a decreasing wall-clock time at a fairly constant average power results in decreasing energy consumption.

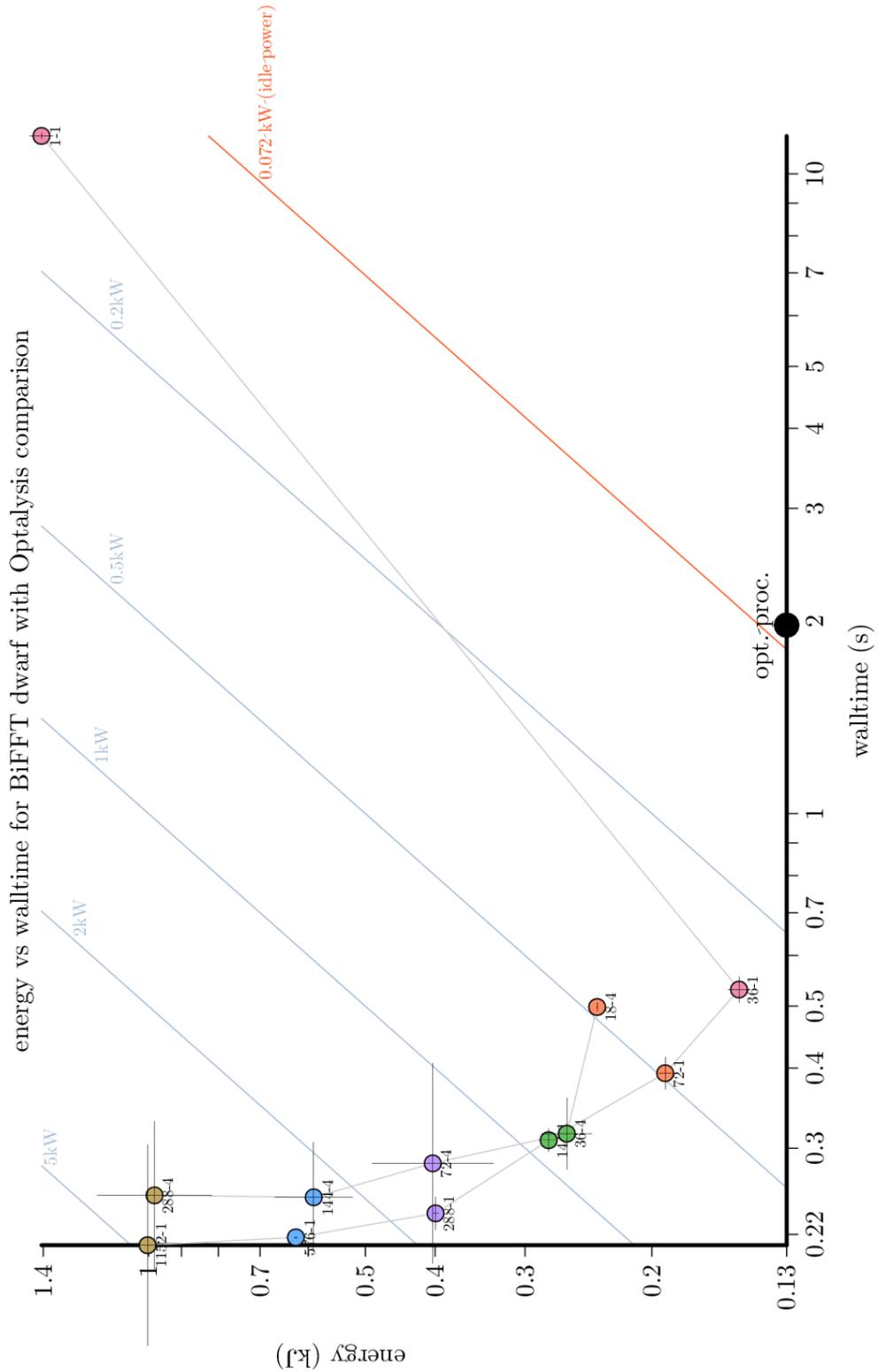

*Figure 2: Log-log plot of the energy consumption vs. wall-clock time for the BiFFT dwarf and corresponding to the combination of one direct and one inverse transformation for 525 fields. Each data point is the result of averaging the outcome of two separate runs and the black crosses represent error bars. Grey lines connect runs with the same number of OMP threads (1 resp. 4). Same-coloured data points use the same number of nodes. Added are lines of constant power (light blue lines), including the power delivered by a node in the idle state (orange line). Indices below each data point denote the number of MPI tasks (left number) as well as the number of OMP threads per task (right number). The black dot represents the estimate of the Optalysys optical processors (see section 5.5).*

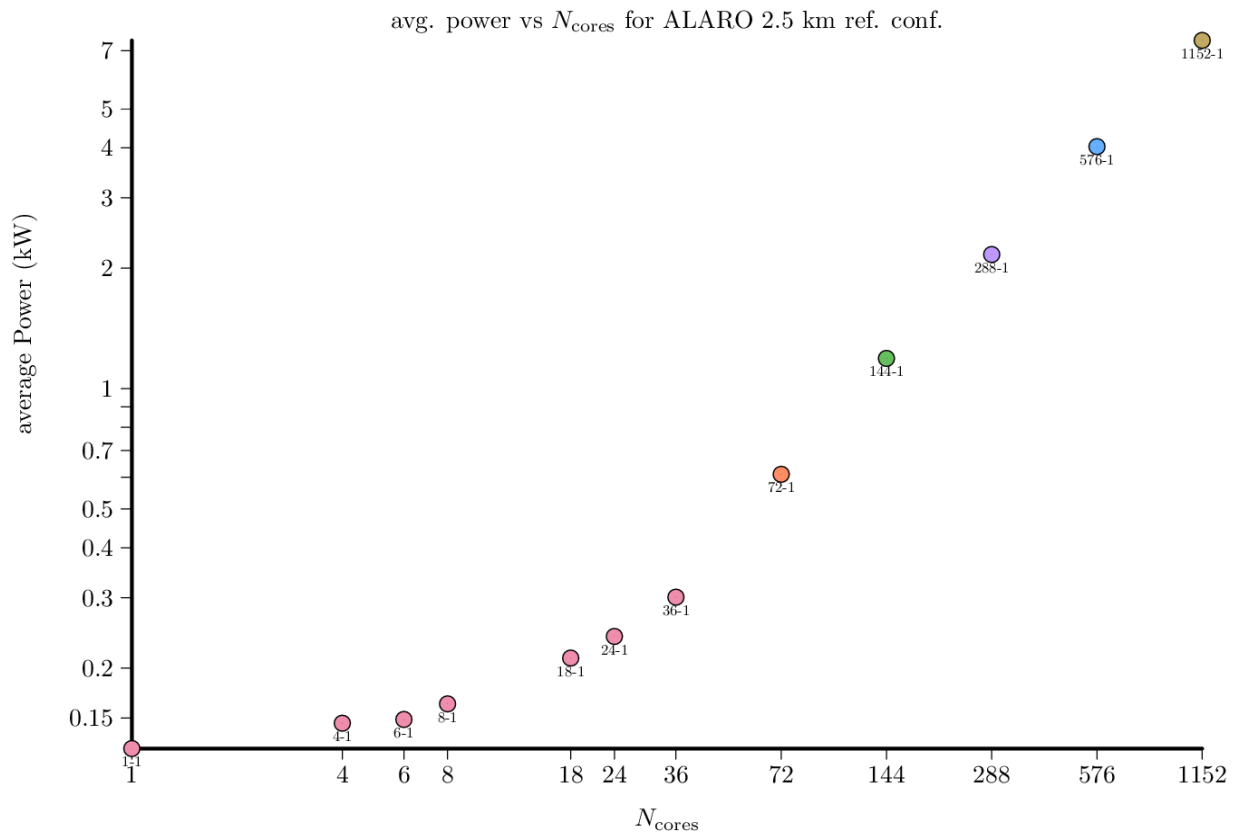

*Figure 3: Log-log plot of the energy consumption vs walltime for the Acraneb dwarf. Only pure MPI jobs were simulated. The colours and added lines have the same meaning as in Figure 2.*

Subsequently, as the number N of used cores/nodes increase, the average power becomes more linearly dependent on N. For a perfectly scaling application (i.e. $t_N = t_1/N$), the consumed energy curve becomes flatter. Due to the fact that real applications are not perfectly scalable, their wall-clock times gradually reach a lower limit and the energy consumption then tends to increases linearly with N. The resulting energy vs. wall-clock time curve has a U-shape and reaches a local minimum at a certain value of N. This does not imply that the local minimum is to be associated with any notion of an optimum configuration that is universal, depending on whether the user needs less energy consumption or less wall-clock time.

It should be noted, however, that the wall-clock time scaling of our pseudo-parallel simulations here do not scale as well as the more representative results reported in D4.3. The latter were taken directly from internal ALADIN profiling data performed during runs of the ALARO 2.5 km reference configuration. This mismatch is probably due to the fact that for increasing $N_{cores}$, the effective workload per core in this experimental setup decreases to such an extent that the initial setup becomes the dominant factor for the wall-clock time. So in reality, we expect the curve to be stretched out to lower wall-clock times.

### 4.4 ALARO 2.5 km reference configuration

The results reported here represent energy measurement runs performed with the ALARO 2.5 km reference configuration (see D4.3) on `cca`. These are pure MPI runs (i.e no OMP threads) of a 12h forecast on a 540 x 450 grid and 65 vertical levels. In order to get a better view of the total curve, we also include runs which do not fill an entire node. For the other, multi-node runs we include pie charts showing the relative energy consumption for the BiFFT and Acraneb dwarfs. Since we cannot directly measure these contributions during the ALARO runs, we proceeded as follows. From the corresponding individual dwarf runs (i.e. same parallel setup), we computed the average power consumption per run. Then, from the measured wall-clock time contribution of each dwarf in the corresponding ALARO run, one can multiply that with the corresponding average power and obtain an estimate of the energy contribution of the dwarf to the ALARO run.

Figure 4 shows the consumed energy vs. wall-clock time measurements for the ALARO 2.5 km reference configuration. Similarly to the BiFFT and Acraneb cases, the curve has a U-shape.

The pie charts indicate that the relative contribution to the energy consumption of both the BiFFT and Acraneb dwarfs decrease as the number of cores increases. The contribution of the Acraneb dwarf remains fairly constant, just as its relative contribution to the total wall-clock time did (see D4.3). On the other hand, the behaviour of the BiFFT dwarf is different. Its relative contribution to the wall-clock time increases significantly with increasing $N_{cores}$, because of the ever increasing amount of communications required. This does not affect the energy measurements however, since the communications are not taken into account during the energy measurements. In this sense, the energy pie charts may be a better representation of the computational workloads than the wall-clock time pie charts are.

Figure 5 shows the average power during runs as a function of number of used cores. It is seen that when less than one node is being used ($N_{cores} < 36$ in our case), the average power increases slowly with $N_{cores}$ initially and becomes more linearly dependent on $N_{cores}$ once full nodes are used. (The graph of course is a log-log graph but the slope of the linear part is very close to 1).

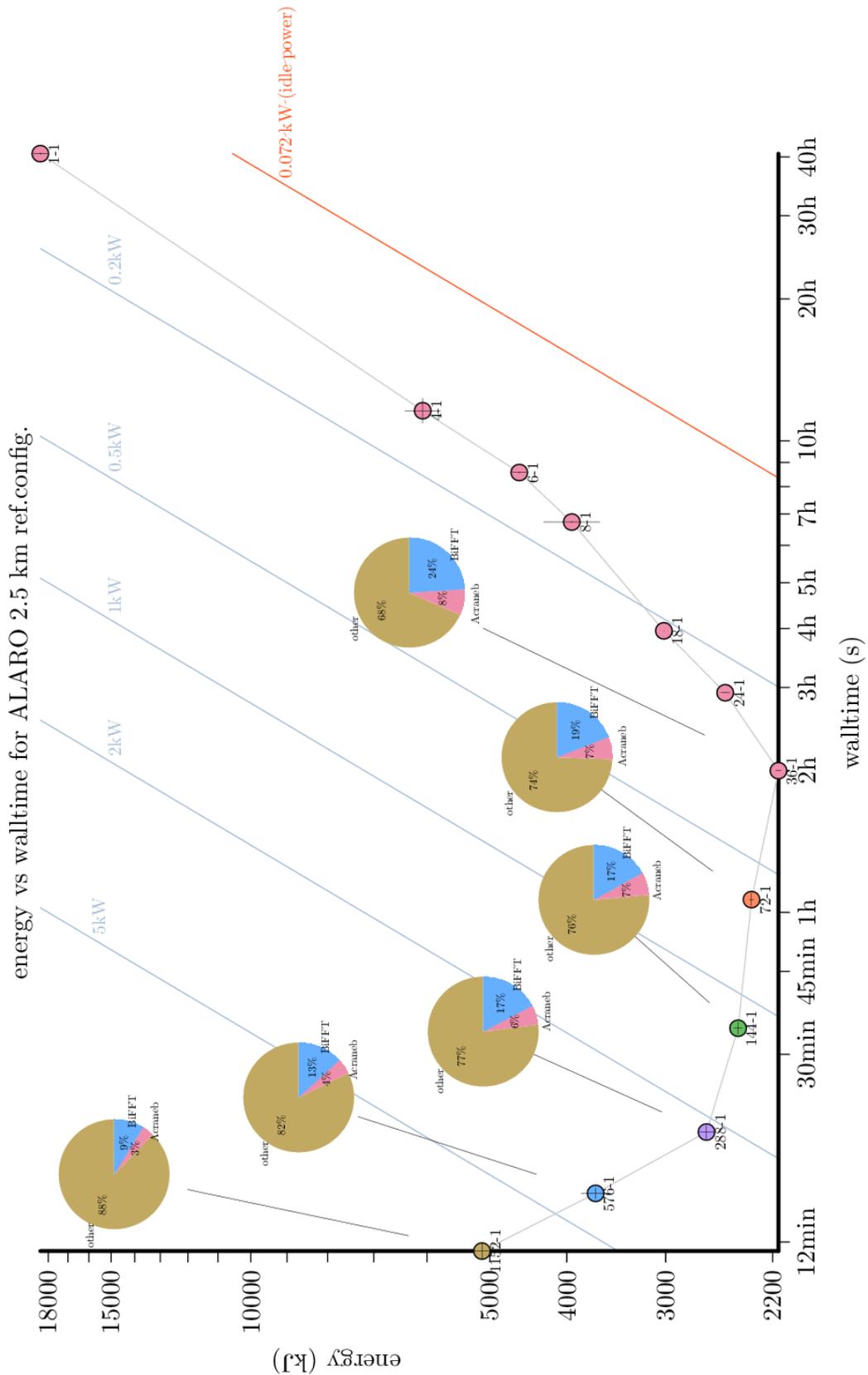

*Figure 4: Log-log plot of the energy consumption vs. wall-clock time for the ALARO 2.5 km reference configuration. Only pure MPI jobs were simulated. The colours of the data points and added lines have the same meaning as in Figure 2. The pie charts are estimates of the relative energy contributions of the BiFFT and Acraneb dwarfs for full-node runs. Thicker grey lines connect the pie charts to the corresponding data point.*

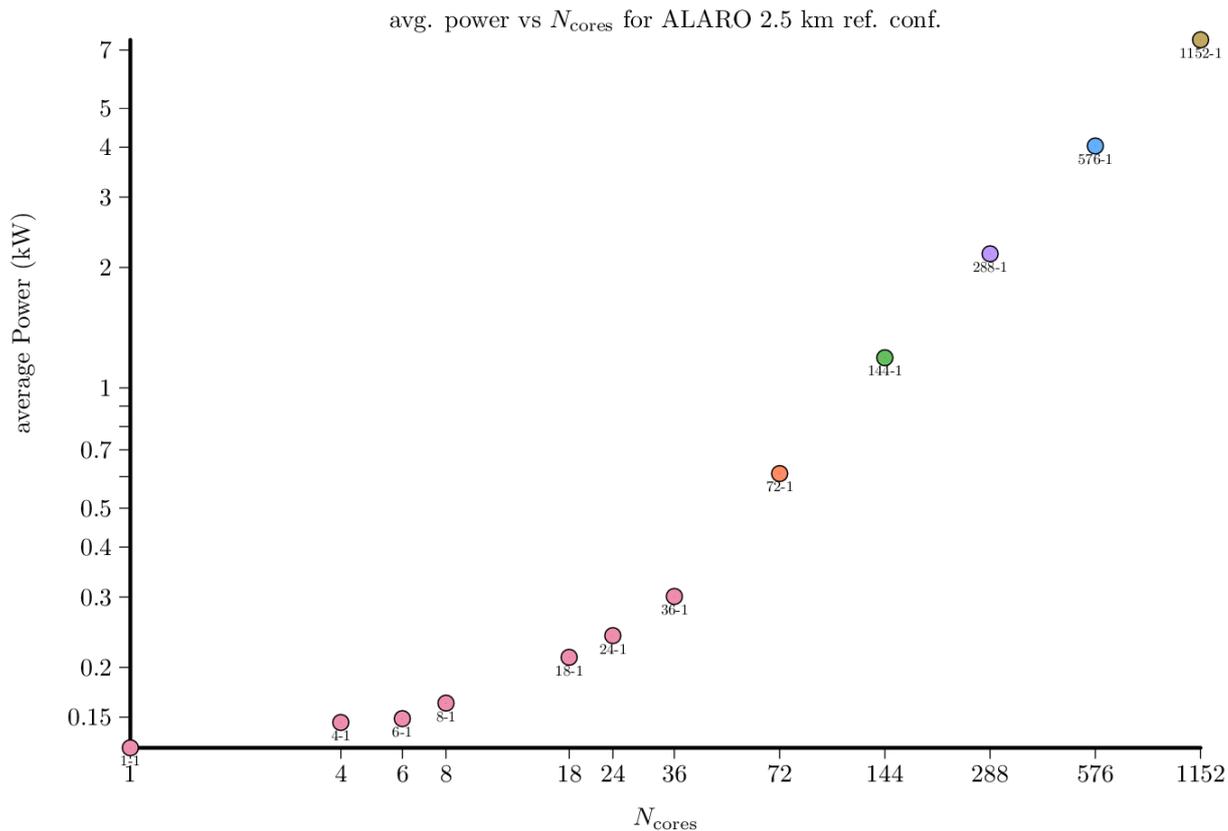

Figure 5: Log-log plot of the average power vs. $N_{cores}$ as measured for the ALARO 2.5km reference configuration. The colours of the data points and added lines have the same meaning as in Figure 2.

### 4.5 COSMO-EULAG reference configuration

The COSMO model which implements the 2.2km reference configuration uses the pre-operational EULAG dynamical core. The model also embeds 2 of the ESCAPE dwarfs; the MPDATA advection and GCR Krylov elliptic solver dwarfs. The model is being run on a 520x350 horizontal grid, using 60 vertical levels. The setup is based on the (former) operational setup of MeteoSwiss, forecasting atmospheric flow over the Alps. Parallelization employs pure MPI two-dimensional domain decomposition and all 36 physical cores per `cca` node are used. The integration time step is set to 20 seconds. Due to the fully explicit, un-split three-dimensional formulation of conservative MPDATA advection, the time-stepping is restricted predominantly by the vertical Courant number. Whenever necessary (i.e. 3D Courant number exceeding unity), the dynamical core is sub-stepped within the physics time step, performing two or more dynamical sub-steps while assuming constant right-hand side forcings stemming from the physical parameterizations. For the meteorological situation at hand, this leads to about 1.94 iterations of dynamics per time step of the model physics.

Figure 6 shows the results of the energy measurements as a function of wall-clock time. There are only 3 data points because on the one hand, due to memory size restrictions, COSMO-EULAG runs could not be performed on setups with 72 cores or less. On the other hand, the run using 1152 cores crashed because of issues with the Courant number treatment. The comparison with the ALARO 2.5 km reference configuration is not straightforward because of non-identical grids, different number of vertical levels, different time step strategies etc., but COSMO-EULAG seems to be more energy demanding. Runs that use the same time step would probably produce a more fair comparison.

The pie charts indicate that the relative energy contribution of the 2 mentioned dwarfs remains fairly constant over all runs. In relation to their relative wall-clock time contributions (see Figure 12 in D4.3) their percentage of energy consumption is slightly higher compared to the rest of the code, indicating that they operate at somewhat higher power than the non-dwarf parts.

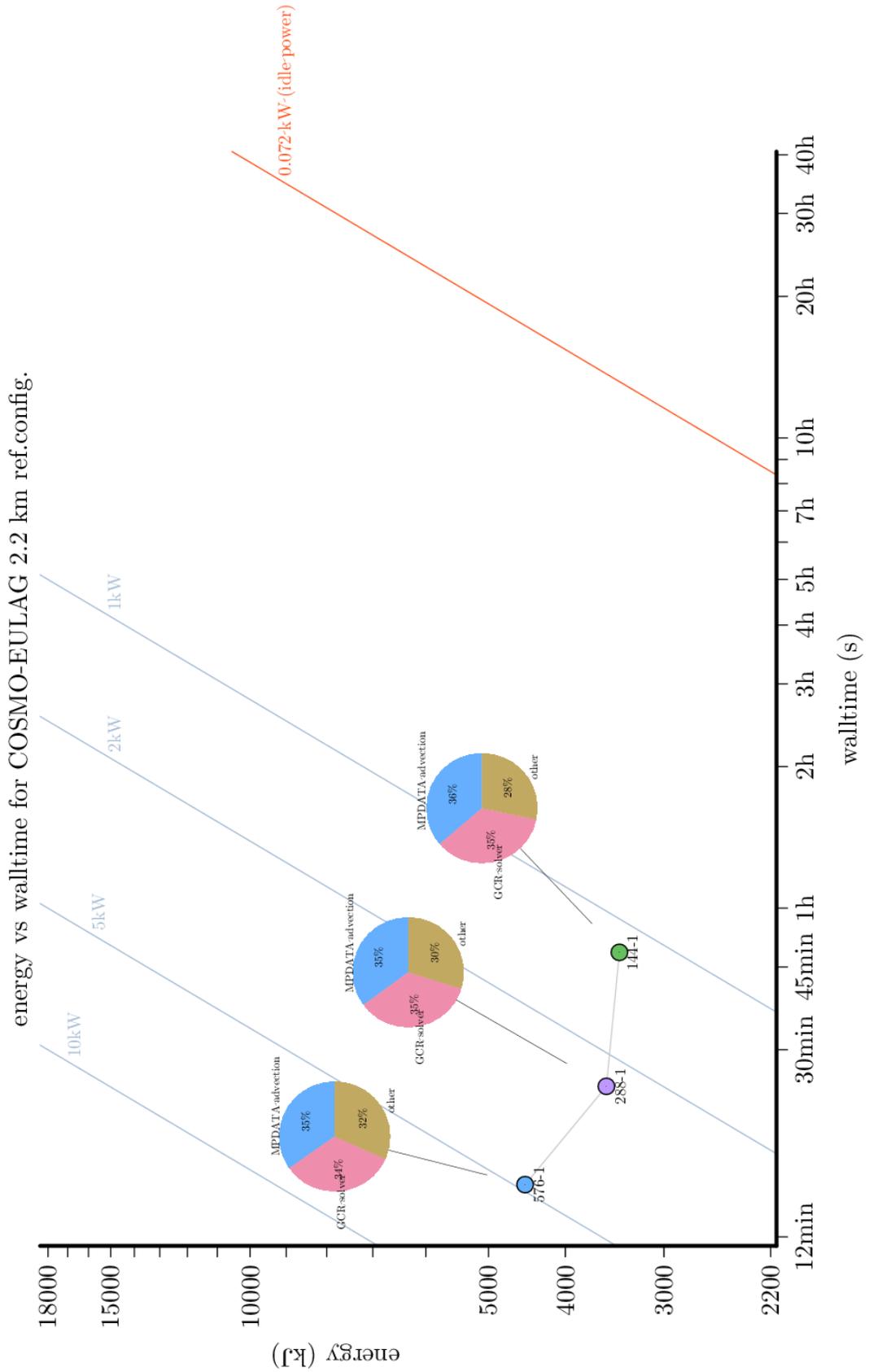

Figure 6: Log-log plot of the energy consumption vs. wall-clock time for the COSMO-EULAG 2.2 km reference configuration. Only pure MPI jobs were simulated. The colours of the data points and added lines have the same meaning as in Figure 2. The pie charts are estimates of the relative energy contributions of the MPDATA and GCR elliptic solver dwarfs for full-node runs.

# 5 Optalysys FTX Power Measurement

## 5.1 Introduction

Optalysys are developing high-performance optical coprocessors. As well as improving the optical capability of such systems, a key aspect is the power, flexibility, and interface to the optics. Hence, Optalysys have undertaken the development of a custom hardware drive solution to transfer data into the optical domain – using spatial light modulators (SLMs) – and back into the electronic domain – using high-speed, high resolution cameras.

This hardware has the potential to support a number of different optical applications, by using different configurations of SLMs in different optical systems. The purpose of this work is to document the power consumption of this system, and then determine the power efficiency in different systems performing operations relevant to numerical weather prediction (NWP).

## 5.2 Background

Coherent optical processing has a prodigious history. A rich variety of optical processing systems have been conceived and developed, and Optalysys have continued this work with systems specific to the requirements of NWP forecasting systems. The specific operations and optical architectures considered are documented in D3.3 section 4.4. .

However, a critical aspect to realizing the potential of optical processing systems is the interface to a traditional computing platform.  Bridging this gap has been a significant undertaking for Optalysys, and has resulted in the development of a custom PCIe drive board (see Figure 7).

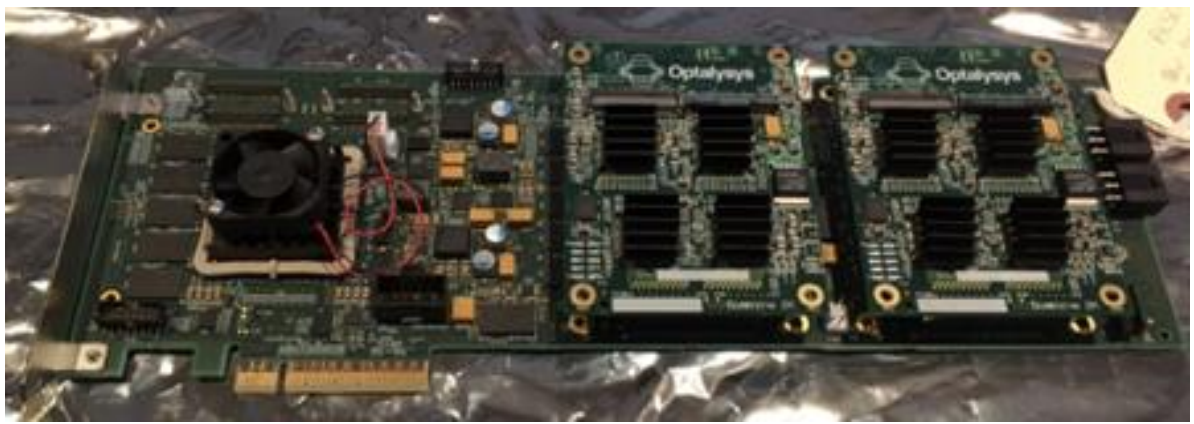

*Figure 7: The Optalysys FTX PCIe drive electronics.*

This board interfaces to a host machine over PCIe and has direct memory access (DMA) to the system memory (RAM). It provides an interface to 4 SLMs and 2 cameras.

The cameras are 4K (4096x3072). Initially, they operate at 100 Hz, but a future firmware upgrade will unlock 300 Hz operation, and 600 Hz half-frame operation, dramatically increasing the potential data throughput.

The SLMs are driven in pairs by two daughter-boards. Interchangeable daughter-boards can be used to drive different SLMs. Currently, there are two options:

- High speed binary SLMs, (Forth Dimension Displays QXGA). These operate at 2.4 kHz, QXGA resolution (1536x2048). They are well-suited to provide high input bandwidth to the optical system.
- Greyscale SLMs (Sony SXRD). These operate at 120 Hz, HD resolution (1920x1080). These are well-suited to use as optical filters to modulate the optical information.

Note that the effective performance of the system is given by the highest framerate component; the lower framerates of other components reduce the flexibility, but not the roofline performance.

The overall electronic configuration of the system is shown in Figure 8.

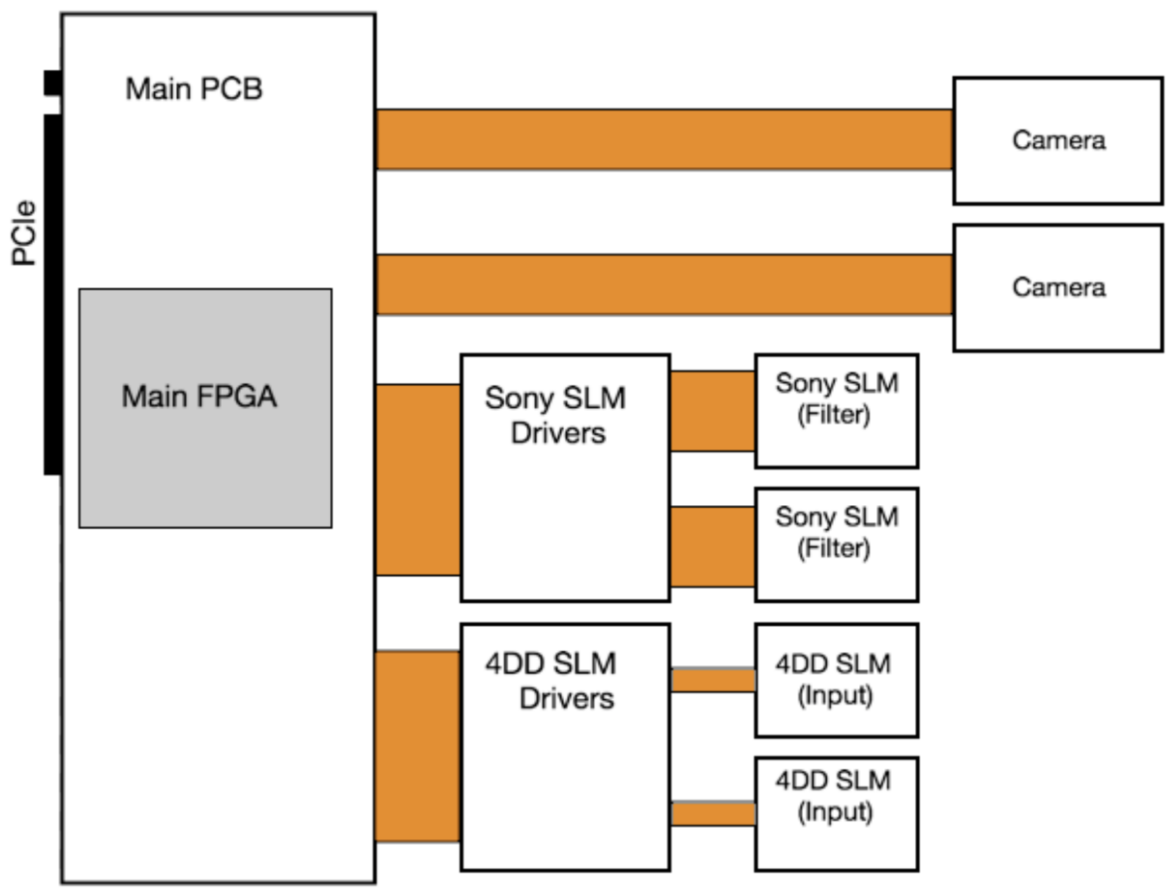

*Figure 8: A block diagram showing the hierarchy of the different electronic drive components.*

Not shown is the laser diode, which is also powered and driven by the PCIe board.

### 5.3 Method

In order to assess the efficiency of the optical approach, we need to measure two things:

1. The power draw of the electronic hardware.
2. The rate of operations performed.

#### 5.3.1 Measuring the power draw

In order to measure the power draw of the electronics, the power delivered to the entire optoelectronic system was measured using a Picotech current clamp around the PCIe power-delivery line. The design of the device is such that power is not drawn over the PCIe connector; the current through the power supply lines constitutes the entire power draw. The current was recorded using an oscilloscope and stored on a connected PC. The instantaneous and mean power are then deduced using *Power = Current x Voltage*.

A typical oscilloscope trace, showing the variation in the current draw, is shown in Figure 9. There are small, local, temporal variations, so we measure the average power draw in this work with the system operating continuously.

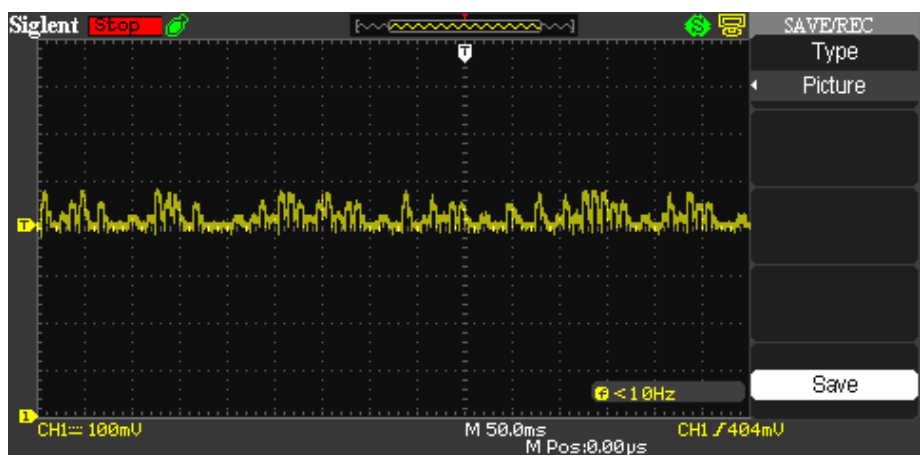

*Figure 9: An oscilloscope trace of the current drawn by the PCIe board, as measured by a current clamp (0.1V/A).*

A typical profile of the power use of the device is shown in Figure 10. The passive power consumption is shown at 36W immediately on power-up of the host PC. Loading the drive results in a further increase in power draw. The main figure we are interested in is the power draw when the application is actually running. Note, that at the end of the run the power consumption does not revert to the inactive level. This is an issue that should be dealt with in further iterations of the driver with improved resource deallocation.

The relevant number is the power drawn during the actual execution of the application itself. This is the number we will use going forwards.

The specific hardware we are investigating is the case when one of the pairs of SLMs is the fast binary 4DD devices, and one pair is the slower, greyscale Sony devices. The power consumption figure is shown in Table 1. Note that the result is greater than the peak value shown in Figure 10 as it also includes the power for the fast SLM displays, which is delivered separately.

| System | Average Power draw |
| --- | --- |
| 2 4DD, 2 Sony, 2 camera | 66 W |

*Table 1: The average power draw in different scenarios.*

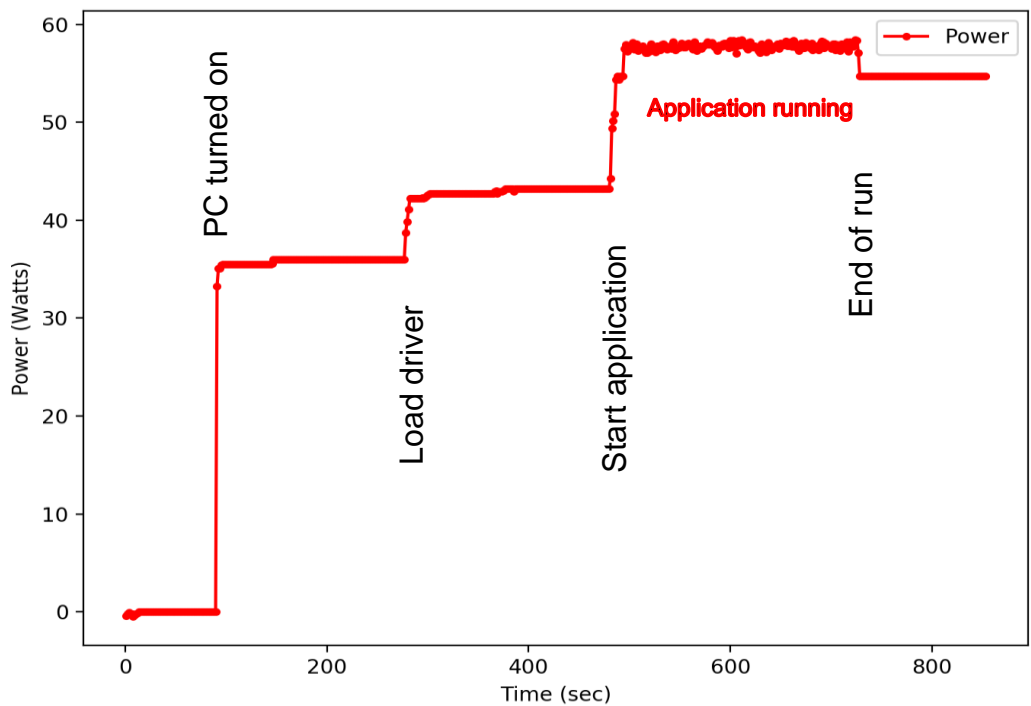

*Figure 10: Power consumption of the device from power on of the host PC, to driver loading, through running a benchmarking application.*

### 5.3.2 Rate of operations performed

The specific optical setup determines the effective operations which have been performed for comparison to conventional platforms. Some configurations make efficient use of the optics, while others use the optical elements less efficiently in order to effectively emulate a specific digital operation.

Essentially, if we are considering a scenario where we are digitally emulating a specific optical system, performance is in favour of the optics. Conversely, if we are considering a scenario where the optics is emulating a specific digital operation, performance is in favour of the conventional platform.

We will consider the following different system scenarios, which are in order of increasing amenability to to capitalize on the underlying performance of the optics.

1. A system to perform 2D Fourier transforms (the BiFFT dwarf).
2. A system to perform spherical harmonic transforms (the sphericalHarmonics dwarf)
3. A system to perform optical convolutions, for pattern matching or deep learning applications.

#### 5.3.2.1 BiFFT dwarf

This dwarf is in some sense the simplest to implement optically: a simple lens performs a 2D Fourier transform. However, as is fully discussed in Deliverable D2-1, in fact it is not straightforward to encode appropriate complex data into the optical domain, nor read it back. In order to effectively encode the data, SLMs have to be used in combination to achieve appropriate modulation; and due to the fact that the camera detects intensity rather than complex amplitude, multiple measurements are required. Finally, the slower, lower-resolution greyscale SLMs must be used.

The most logical way to use the hardware is with exclusively greyscale SLMs, paired up to give the appropriate input modulation, and each associated with one camera. This means we have sufficient hardware for effectively two systems, which can be used to perform the two measurements required to assemble, say, a Complex-to-Real transform, or in two cycles can perform the four measurements required to perform a Real-to-Complex transform.

Hence, the system performance using this hardware would be:

- 1 Complex-to-Real 1920x1080 2D FT in 10ms.
- 1 Real-to-Complex 1920x1080 2D FTs in 2x10ms = 20ms.

Note that these resolutions and performance have not been demonstrated, and there are still engineering hurdles to overcome, but they are in principle possible with this hardware. However, the performance will not be very competitive due to being limited by the slow refresh rate of the Sony SLMs.

Lower resolution operations can be performed, but with no performance benefit.

#### 5.3.2.2 Spherical harmonics transform

An optical implementation of the spherical harmonic transform is presented in Deliverable D3-3. It involves using an optical correlator architecture to extract coefficients of the spherical harmonic transform. In particularly, we will consider the multichannel astigatic optical processor architecture, which performs all the requisite Fourier transforms optically, then extracts Legendre coefficients using correlation filters. We will assume that it is possible to make use of fast binary filters and that we can spread the results across the camera sensor to compensate for the camera being slower than the filter SLM, although both of these features require further research.

Assuming this is the case, we can use 1920x1080 input data, and extract an entire set of Fourier coefficients for a given Legendre coefficient at a rate of 2400 Hz (0.42ms per frame). Again, we will have to make use of multiple measurements to overcome the fact that the camera detects intensity.

A pair of greyscale input SLMs will be required to overcome the modulation limitations, along with a fast binary filter SLM. Hence, this system would unfortunately not make good use of the drive ability of the electronics, leaving a surplus fast binary SLM and camera. The overall system performance is:

From a 1920x1080 resolution input, extract one set of Legendre coefficients (across all Fourier coefficients) in 4x0.42ms = 1.7ms.

#### 5.3.2.3  2D convolutions

This is the application to which this hardware is most suited, and where it can provide truly competitive performance. By making use of two fast binary input SLMs, two greyscale filter SLMs, and the two cameras, the system is capable of performing

- 2 2048x1536 2D convolutions in 0.42ms.

That is, the equivalent digital system would have to perform

- 4 2048x1536 complex 2D FTs in 0.42ms (plus a negligible element-wise matrix multiplication)

While the operation of the optical system is constrained (it cannot perform *arbitrary* convolutions), this is the equivalent computational power required to emulate the optics. It is in this correlation application that, not only is the optics competitive, but the low-precision 'noisy' nature of the optics is not a significant issue. While high precision is important in NWP, it is less important in pattern-matching and data-processing applications.

### 5.4 Results

We can summarize these applications in the Table 2.

| Operation | Time (ms) | Energy consumption (mJ) |
|---|---|---|
| 1920x1080 C2R 2D FT | 10 | 660 |
| 1920x1080 R2C 2D FT | 20 | 1320 |
| 1920x1080 Legendre coefficient extraction | 1.7 | 112.2 |
| 2048x1536 C2C 2D FT[1] | 0.1 | 6.6 |

*Table 2: Energy consumption of the various Optalysys hardware.*

The significant increases in energy per operation are clear, as the properties of the operation become more restricted and aligned with the inherent properties of the optics.

In general, it is clear that when the optics are made to perform a specific task pertinent to NWP, it suffers in terms of the effective performance and efficiency. However, the latent performance of the optics is significant. The challenge is finding an application which fits naturally with the ability of the optics. Search applications and, potentially, convolutional neural nets appear to be appropriate applications for the system.

### 5.5 Comparison with BiFFT energy measurements on CPU

From the measurements shown in section 5.3.1 we can make a comparison of wall-clock time and energy consumption for the BiFFT dwarf between the Cray XC40 cluster and the Optalysys optical processors. The forward respective inverse BiFFT transformations correspond to real->complex, respective complex->real transformations on the optical processors. This means that one forward combined with one inverse transformation takes 10ms + 20ms = 30ms.

However, the HD resolution of the Greyscale SLMs (1920x1080 pixels) could be taken advantage of more optimally by configuring the data input so that a maximum of 8 separate transformations on the grid of the BiFFT dwarf (540x450 gridpoints) are handled simultaneously. Therefore, one real to complex transform combined with one complex to real transform on the optical processors can account for the combination of 8 forward and 8 inverse transf. for 1 field on the grid of the BiFFT dwarf, while requiring a wall-clock time of 30ms.

---

[1] In the context of 2D convolutions

Recall that Figure 2 showed wall-clock time and energy consumption of one forward plus one inverse transformation for 525 fields, so the corresponding numbers for the optical processors then become:

$$\Delta t = 525 * \frac{30\ ms}{8} = 1.97\ s$$

$$\Delta E = 66\ W * 1.97\ s = 130\ J$$

In Figure 2, this estimate is represented as the black dot. Compared to the Cray XC40 cluster, the optical processors are fairly slow (only being faster than the serial runs on `cca`) but they consume the least amount of energy. As a matter of fact, the optical processors are operating at a power that is just below the idle power of a compute node on the Cray XC40 cluster.

As already discussed in D2.1, the optical processors are not competitive with traditional CPU's at current NWP grid resolutions in terms of wall-clock time, a conclusion made worse by the inferior numerical precision achieved. One can expect the former issue to become less severe if NWP grids keep increasing in resolution. Due to the $O(1)$ wall-clock time dependency for optical processors compared to the $O(n * \log n)$) dependency for CPUs (plus communications), a critical resolution will be reached at some point where the advantage turns in favour of the optics. However, given the low numerical accuracy, the usefulness of optical processors for NWP remains in doubt.

## 6   Conclusion

In this deliverable we have reported on energy consumption measurements of a number of NWP models/dwarfs on the Intel E5-2697v4 processor. The chosen energy metrics and energy measurement methods were documented. Energy measurements were performed on the Bi-Fourier dwarf, the Acraneb dwarf, the ALARO 2.5 km reference configuration as well as on the COSMO-EULAG reference configuration. The results showed a U-shape dependence of the consumed energy on the wall-clock time performance. This shape was explained from the dependence of the average power of the compute nodes on the total number of cores used. Even if such a shape has a local energy minimum, it does not imply a "one size fits all" best solution since not everyone has the same requirements and constraints.

The relative energy consumption contributions of the BiFFT dwarf to the ALARO reference configuration showed a different behaviour compared to its relative contribution to the wall-clock time. While communications cause a significant increase to the relative wall-clock time with increasing number of cores, the same does not happen to its relative energy consumption contribution, simply because the energy consumption of the communications are not accounted for. The energy consumption pie charts are therefore more representative of the relative computational workloads of the dwarfs.

A comparison between the ALARO 2.5 km and COSMO-EULAG 2.2 km reference configuration indicates that the latter consumes more energy and requires slightly more runtime. However, making a fair conclusion is difficult because of slightly differing setups (especially concerning the used time step).

We compared the energy consumption of the BiFFT dwarf on the E5-2697v4 processor to that on the Optalysys optical processors. The latter were found to be much less energy costly, but it is also the only metric where they outperform the classical CPU. They are non-competitive as far as wall-clock time and especially numerical precision are concerned. However, it is clear that the BiFFT application is not amenable to the optical approach, as the specific requirements of the operation lead to sacrificing a lot of the potential performance. Other applications present a more appropriate application of this technology.

Finally, we suggested an interesting candidate energy metric called the *energy roofline*. It is an energy analogue to the well-known time-based performance roofline. It allows an assessment of whether a piece of code is compute-bound or memory-bound in terms of energy consumption. Such a tool could be useful for more detailed future energy optimization studies for NWP models/dwarfs.

The measurements described here were not meant to be exhaustive (e.g. no GPU measurements) and only represent an initial step in the process of energy consumption studies of numerical weather prediction models/dwarfs. Future tests may consider many more aspects and variables that can affect energy consumption such as other CPU architectures (since energy varies with different socket/cpu design), DRAM etc. But also parts of a code (e.g. loops or individual algorithms) can be studied through the use of the energy roofline model.

weather dwarfs. 30th International Conference on Parallel Computational Fluid Dynamics.

## Document History

| Version | Author(s) | Date | Changes |
|---|---|---|---|
| 0.1 | Joris Van Bever, Alex McFaden, Zbigniew Piotrowski, Daan Degrauwe | 9/5/2018 | original |
| 1.0 | Joris Van Bever, Alex McFaden | 28/5/2018 | Internal review corrections/suggestions |
| | | | |
| | | | |

## Internal Review History

| Internal Reviewers | Date | Comments |
|---|---|---|
| Mike Gillard (LU) | 17/05/2018 | Approved with comments |
| Michal Kulczewski (PSNC) | 18/05/2018 | Approved with comments |
| | | |
| | | |

## Effort Contributions per Partner

| Partner | Efforts |
|---|---|
| RMI | 11.08 |
| OSYS | 1 |
| PSNC | 0.75 |
| **Total** | **12.83** |

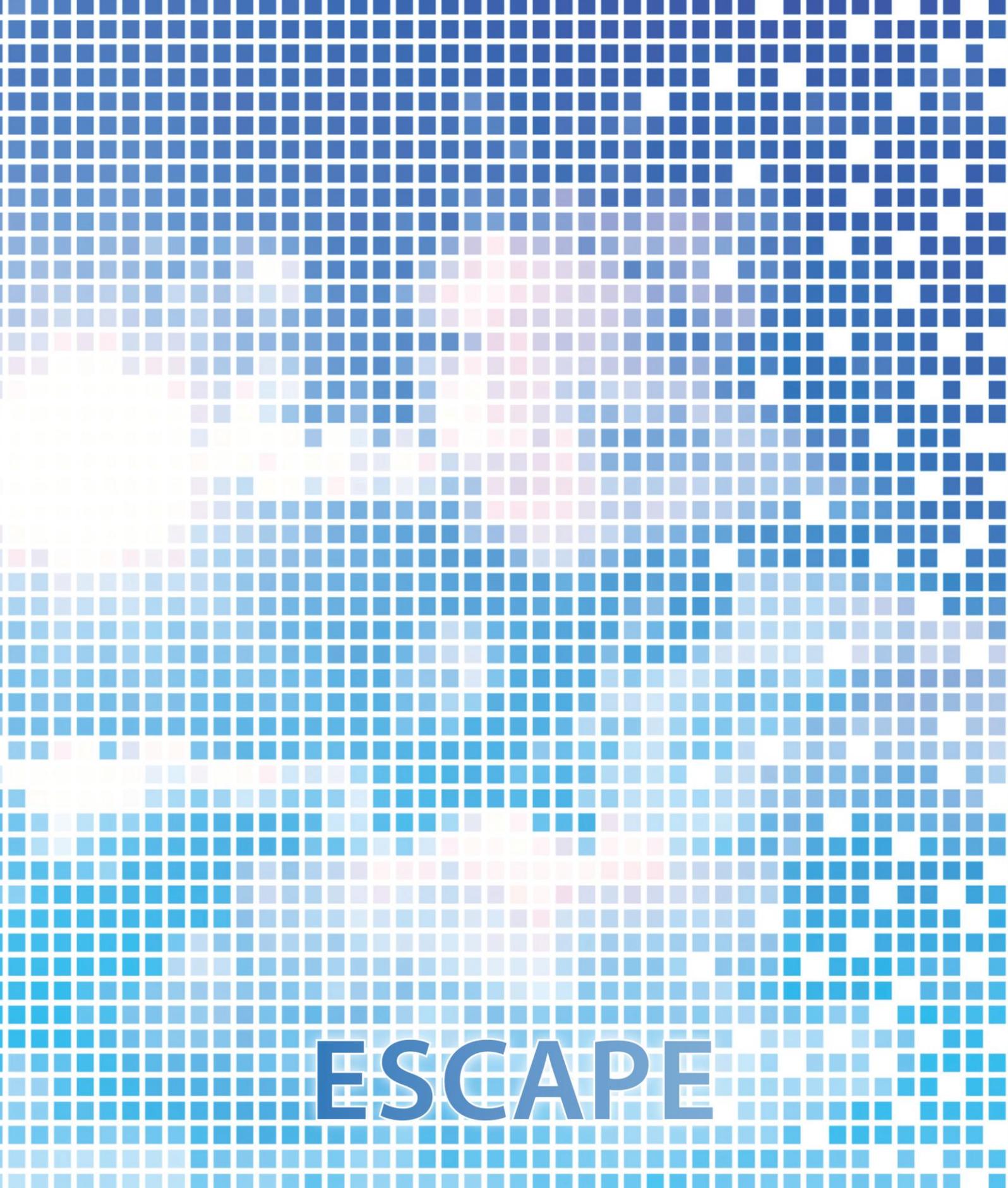